\documentclass[amsmath,amssymb,prl,aps,showpacs,twocolumn,10pt,superscriptaddress]{revtex4-1}
\usepackage{amsmath,amsfonts,amssymb,amsthm,graphics,graphicx,epsfig,bbm}
\usepackage[colorlinks=true,citecolor=blue,linkcolor=red,urlcolor=blue]{hyperref}
\usepackage[usenames]{color}
\usepackage{graphicx}
\usepackage{subfigure}
\usepackage{amsmath}
\usepackage{epsfig}
\usepackage{dcolumn}
\usepackage{bm}
\usepackage{color}
\usepackage{epstopdf}
\usepackage{amssymb}
\usepackage{amstext}
\usepackage{latexsym}
\usepackage{hyperref}
\usepackage{amsfonts}
\usepackage{psfrag}
\usepackage{xcolor}
\usepackage[normalem]{ulem}
\usepackage{dsfont}
\usepackage{txfonts}
\usepackage{overpic}
\usepackage{ifthen}
\usepackage{comment}

\newcommand{\ket}[1]{\vert #1 \rangle}

\newcommand{\Tr}{\mathrm{Tr}}

\newcommand{\an}[2]{\ifthenelse{\equal{#1}{}}{\ensuremath{\hat{#1}_{#2}}}{\ensuremath{\hat{#1}^{\protect\phantom{\dagger}}_{#2}}}}

\newcommand{\blue}[1]{\textcolor{blue}{#1}}

\begin{document}
\title{Chimera Time-Crystalline order in quantum spin networks}

\author{A. Sakurai}
\email{akitada-phy@nii.ac.jp}
\affiliation{National Institute of Informatics, 2-1-2 Hitotsubashi, Chiyoda-ku, Tokyo 101-8430, Japan}
\affiliation{School of Multidisciplinary Sciences, Department of Informatics, SOKENDAI (The Graduate University for Advanced Studies), Shonan Village, Hayama, Kanagawa 240-0193, Japan}

\author{V. M. Bastidas}
\email{victor.m.bastidas.v.yr@hco.ntt.co.jp}
\affiliation{NTT Basic Research Laboratories \& Research Center for Theoretical Quantum Physics,  3-1 Morinosato-Wakamiya, Atsugi, Kanagawa, 243-0198, Japan} 
\affiliation{National Institute of Informatics, 2-1-2 Hitotsubashi, Chiyoda-ku, Tokyo 101-8430, Japan}

\author{W.~J.~Munro}
\affiliation{NTT Basic Research Laboratories \& Research Center for Theoretical Quantum Physics,  3-1 Morinosato-Wakamiya, Atsugi, Kanagawa, 243-0198, Japan} 
\affiliation{National Institute of Informatics, 2-1-2 Hitotsubashi, Chiyoda-ku, Tokyo 101-8430, Japan}

\author{Kae Nemoto}
\affiliation{National Institute of Informatics, 2-1-2 Hitotsubashi, Chiyoda-ku, Tokyo 101-8430, Japan}

\begin{abstract}
Symmetries are well known to have had a profound role in our understanding of nature and are a critical design concept for the realization of advanced technologies. In fact, many symmetry-broken states associated with different phases of matter appear in a variety of quantum technology applications. Such symmetries are normally broken in spatial dimension, however they can also be broken temporally leading to the concept of discrete time symmetries and their associated crystals. Discrete time crystals (DTCs) are a novel state of matter emerging in periodically-driven quantum systems. Typically, they have been investigated assuming individual control operations with uniform rotation errors across the entire system.  In this work we explore a new paradigm arising from non-uniform rotation errors, where two dramatically different phases of matter coexist in well defined regions of space. We consider a quantum spin network possessing long-range interactions where different driving operations act on different regions of that network. What results from its inherent symmetries is a system where one region is a DTC, while the second is ferromagnetic.  We envision our work to open a new avenue of research on Chimera-like phases of matter where two different phases coexist in space.
\end{abstract}

\maketitle
\date{\today}

Symmetries, while they may be a simple concept, have had a profound effect on many fields of physics and are crucial in understanding many natural phenomena~\cite{landau1981,Yang54,Lowe11} as well as the realization of many advanced technologies. This includes our well-known phases of matter (solids, liquids and gases). Crystals as a solid with a periodic nature are one of the most familiar examples of spatial symmetry breaking~\cite{landau1981,Cooper56,Shahar1997,Higgs1964}. One can also think of the temporal dimension and whether such symmetry breaking occurs there~\cite{Wilczek2012,Bruno2018,Watanabe2015}. In fact, temporal symmetry breaking does occur in periodically-driven nonequilibrium systems and the phase of matter that arises is referred to as discrete time crystals (DTCs)~\cite{Sacha2015, Zakrzewski2018, else2019discrete,Else2016, Khemani2016, berdanier2018, Giergiel2018, guo2020condensed, Guo2013, guo2016synthesizing,Russomano2017,  Pizzi2019, PizziB2019, Silva2019, Lemini2018, Gambetta2019, RieraCampeny2020, Estarellas19}. Recently the existence of DTCs has been demonstrated in trapped ions \cite{zhang17}, nitrogen-vacancy spin impurities~\cite{Choi2017}, nuclear spins in molecules~\cite{Pal2018}, superfluid quantum gases~\cite{Smits2018}, ordered dipolar many-body systems~\cite{Rovny2018} and silicon doped with phosphorus~\cite{o2018dissipative}.

\begin{figure}[t]
\centering
\includegraphics[width=0.45\textwidth]{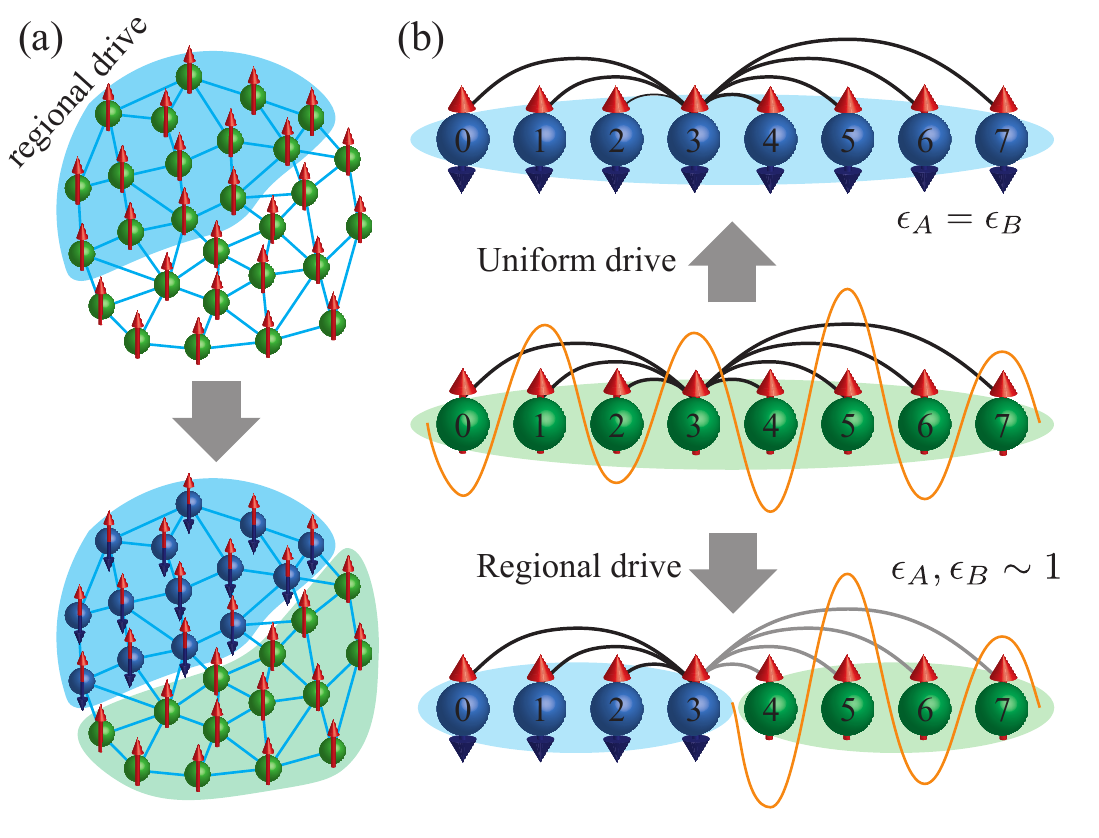}
\caption{ 
Chimera discrete time crystals in a spin network. (a) Schematic illustration of the chimera time crystal in a spin network. The top panel illustrates this network under the effect of regional driving while the bottom panel shows the sublattice $\text{A}$ (blue region), which behaves like a discrete time crystal. In contrast, the region B (green region) is in a ferromagnetic state. (b) Illustrates a $N=8$ site spin chain where we apply regional drivings. A $\pi$-rotation is applied on region A spins $l=0,1,2,3$ while small rotation is applied on the remaining spins that define region B. Here, the black and yellow lines represent the couplings $J^z_{lm}$ between the spins and the disordered potential $W_l$, respectively.  In the case of regional drive, the gray lines represent the weak coupling between the regions A and B. 
}
\label{Fig1a}
\end{figure}

In those recent demonstrations, to generate the DTC one needs to apply individual spin rotations with a uniform error across the entire system~\cite{Zakrzewski2018,Sacha2015,else2019discrete,Else2016,Khemani2016}. Of course there is no reason that one needs to utilize a uniform drive acting on the whole system. Instead the drive could be different for different regions within the system. This is particularly interesting as it means different phases of coexisting matter could be engineered. 
In this letter we investigate the effect of regional driving on a system capable of supporting DTCs and explore its dynamics. We show that multiple phases of matter can coexist within the overall system. Such novel phases of matter are analogous to Chimera states in classical nonlinear systems where synchronized and unsynchronized phases coexist~\cite{kuramoto2002coexistence,Abrams2004,panaggio2015chimera,hagerstrom2012} even in the semiclassical regime~\cite{Bastidas2014}. As such, we are going to consider ``Chimera DTCs" consisting of a DTC and an alternate phase of matter (ferromagnetic). We will generally begin with a DTC in a quantum spin network and then apply a drive to a certain region of that crystal to evolve it into the ferromagnetic phase (alternatively one could start with the ferromagnetic phase for the entire system and apply a drive to a certain region to transform that region into a DTC). This is depicted in Fig.~\ref{Fig1a} (a). Both the original and the new phases of matter coexist at the same time in different regions of space (a Chimera-like state) despite the spin-spin coupling throughout the network.

Let us begin with a $N$-spin quantum network governed by a time-periodic Hamiltonian of the form
 \begin{equation}
         \label{eq:DiscreteTimeCrystal}
  \hat{H}(t)= 
  \begin{cases}
    \hat{H}_1 = \hbar g(1-\epsilon_\text{A}) \sum\limits_{l \in \text{A} }  \sigma_{l}^{x}  +  \hbar g (1-\epsilon_\text{B}) \sum\limits_{l \in \text{B}} \sigma_{l}^{x} & 0 \leq t < T_1 \\
 \hat{H}_2 =    \hbar \sum\limits_{lm}J_{lm}^{z} \sigma_{l}^{z} \sigma_{m}^{z} + \hbar \sum\limits_{l}W_{l}^{z} \sigma_{l}^{z} & T_1 \leq t < T \ ,
  \end{cases}
\end{equation}
with a total period $T=T_1+T_2$ ($\hat{H}_1$ is applied for a time $T_1$ followed by $\hat{H}_2$ for a time $T_2$). The nodes of our network are the individual spins at sites labelled by $l=1,\ldots,N$.  Here $\sigma_{l}^\mu$ with $\mu \in \{x,y,z\}$ are the usual Pauli operators at the $l-$th site. Next $g$ is a drive amplitude chosen such that $g T_1 = \pi/2$.  Our Hamiltonian $\hat{H}_1$ applies separate rotations on the two well-defined spin network regions where we allow for errors $\epsilon_A$ and $\epsilon_B$, respectively. This is shown in Fig.~\ref{Fig1a}~(a) as the blue region A and green region B. 
Further the couplings  $J_{l,m}$  determine the connectivity of the network,  because they can be represented by an edge joining the $l-$th and $m-$th  nodes as depicted in Fig.~\ref{Fig1a}. We also consider the effect of disorder $W_{l}^z \in [-W,W]$ drawn from a uniform distribution with strength $W$. Finally, it is worth mentioning that when $\epsilon_\text{A} = \epsilon_\text{B}\ll 1$ the whole system retains its single DTC nature~\cite{Zakrzewski2018,else2019discrete}.

Now let us evaluate how we can manipulate these phases of matter using regional drives.  We consider the case where the rotation arising from the drive on region A is close to $\pi$ ($\epsilon_A$ is small) while the drive on region B is effectively turned off ($\epsilon_B$ close to one). The local magnetization ${m^z}_{l } (nT)= \langle \sigma_{l}^z (nT)\rangle$ at the $l $-th site measured at stroboscopic times $t_n=nT$ (with $n$ being a natural number) can then be used to monitor the breaking of the discrete time translational symmetry and the emergence of DTC in a given region of the network.

To investigate this stroboscopic behaviour, we employ the Floquet operator for a single drive period~\cite{1883floquet,1998GRIFONI,Polkovnikov2015,2016Restrepo} given by 
\begin{equation}
	\hat{\mathcal{F}} = \exp \Big( -\frac{i}{\hbar} \hat{H}_2 T_2 \Big) \exp \Big( -\frac{i}{\hbar} \hat{H}_1 T_1 \Big)
	\label{eq:FloquetOperatorT}
	\ .
\end{equation} 
The 2T-periodicity of the DTC enables us to unveil the symmetries of the system at times $t_n=2nT$. Here, the stroboscopic dynamics is generated by the square of the Floquet operator $\hat{\mathcal{F}}^2 = \exp\; (-2i \hat{H}_{\epsilon_A,2T}^{\text{eff}} T/\hbar ) $ with $\hat{H}_{\epsilon_A,2T}^{\text{eff}} $ being the systems $2T$-effective Hamiltonian.  Due to $\hat{H}_{\epsilon_A,2T}^{\text{eff}}$ structure  we can use a high-frequency expansion ~\cite{Polkovnikov2015,2016Restrepo} for the driven system to express it in the closed form $\hat{H}_{\epsilon,2T}^{\text{eff}}=\hat{H}_{\text{A}}+\hat{H}_{\text{B}}+\hat{H}_{\text{AB}} $ where
\begin{eqnarray}
\label{eq:Hamiltonians}
\hat{H}_{\text{A}} &=&  \frac{\hbar}{2} \sum_{l,m\in A} J_{lm}^z \sigma_{l}^z \sigma_{m}^z - \frac{\hbar \pi \epsilon_A } {4}\sum_{\substack{l,m \in A }} J_{lm}^z \sigma_{l}^z \sigma_{m}^y
 \nonumber \\
\hat{H}_{\text{B}} &=&  \frac{\hbar}{2} \sum_{l,m \in B}J_{lm }^z \sigma_{l }^z \sigma_{m }^z + \frac{\hbar }{2} \sum_{l\in B} W_{l } \sigma_{l}^z\\
\hat{H}_{\text{AB}} &=& - \frac{\hbar \pi  \epsilon_A }{4T} \sum_{l \in A}  \Big[
	\Big(  \cos{\hat{\theta}_{l }} + 1 \Big) \sigma_{l }^x +
	 \sin{\hat{\theta}_{l}}  \sigma_{l}^y \Big]  \nonumber
\ .
\end{eqnarray}
Here we have assumed for convenience $\epsilon_A \ll1$ and $\epsilon_B=1$ (this simplification will be relaxed in our simulations).

The coupling Hamiltonian $\hat{H}_{\text{AB}}$ is strongly dependent on the operator $ \hat{\theta}_{l}=2W_{l} T_2 + 2T_2\sum_{m\in B} J_{lm}^z \sigma_{m}^z $ for the $l$-th sites in A~\cite{SupplementalInfo}.  
Setting $\epsilon_A=0$, the regions  A and B are decoupled at times $t_n=2nT$.   This indicates that each region preserves the local operators $\sigma^z_l$ and its own symmetries.  Region A holds a $\mathbb{Z}_2$-Ising symmetry $\sigma_{l}^z=-\sigma_{l}^z$ which may be broken in B due to the disorder.  
Remarkably for $\epsilon_A\ll1$, $\hat{H}_{\epsilon_A,2T}^{\text{eff}}$ breaks the $U(1)$ symmetry in region A while  $\hat{H}_{\text{B}}$ associated with region B remains conserved with $[\hat{H}_{\text{B}},\hat{H}_{\epsilon_A,2T}^{\text{eff}}]=[\sigma^z_l,\hat{H}_{\epsilon_A,2T}^{\text{eff}}]=0$. This creates the {\it chimera} DTC where two phases of matter emerges in a network of spins.

\begin{figure}[b]
\centering
\includegraphics[width=0.45\textwidth]{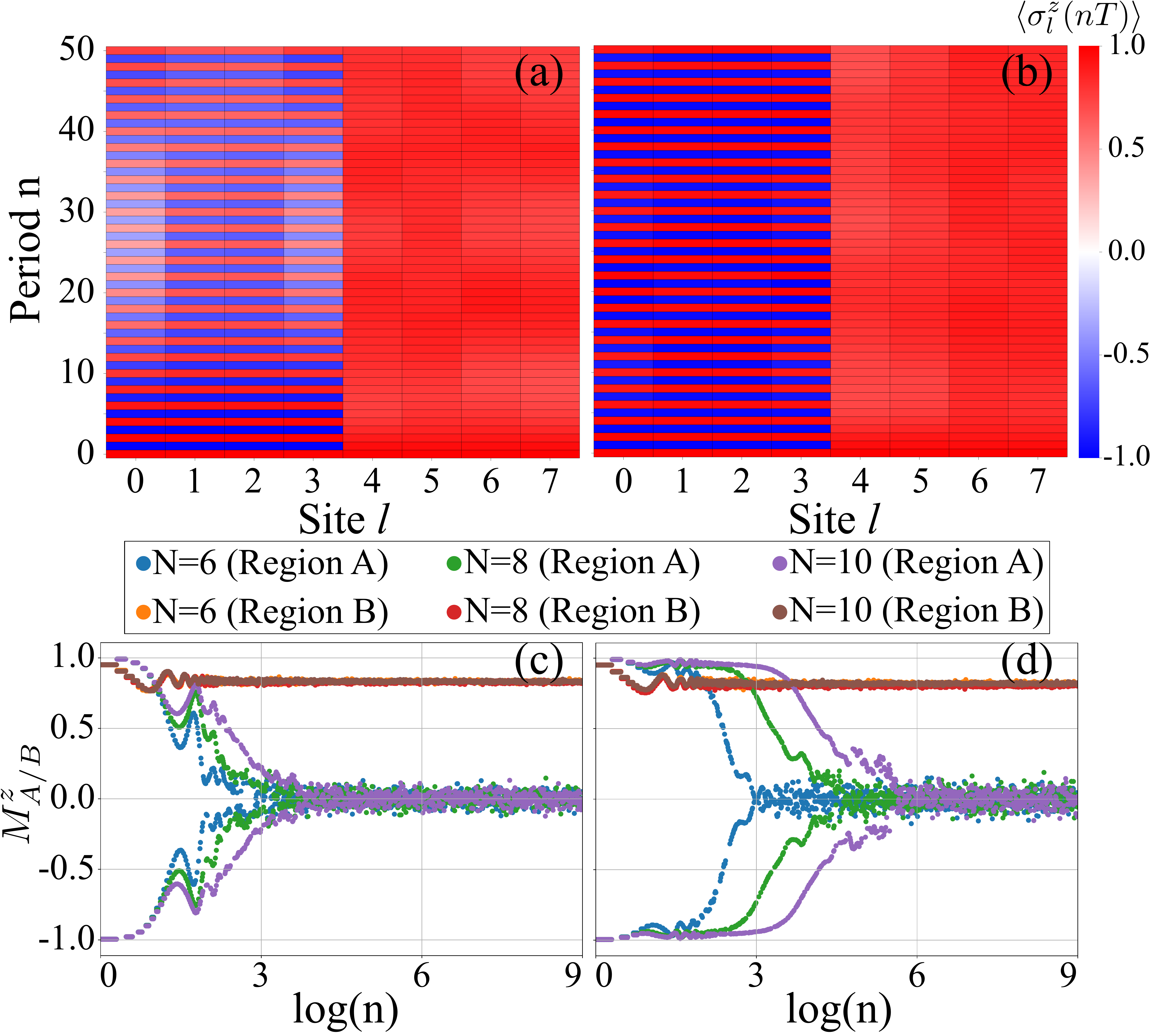}
\caption{ 
The short-time dynamics of the local magnetization $\langle \sigma^z_l (t)\rangle$ for a Chimera DTC in the weak $J_0T = 0.072$ (a) and strong $J_0T = 0.2$ (b) coupling regimes, respectively.  Here, we have chosen $\epsilon_A=0.03$, $\epsilon_B=0.9$, $gT = \pi$, $\alpha=1.51$ and $WT = 2\pi$ with an initial state $|\Psi(0)\rangle_z = |1, 1\cdots1\rangle$. Next the long-time dynamics of the regional magnetization $M^z_{A/B}=2/N\sum_{l\in A/B}\langle \sigma^z_l (t)\rangle$ are shown for  weak (c) and strong (d) couplings and different system sizes $N=6,8,10$.
}
\label{Fig1b}
\end{figure}

To explore the dynamics (the emergence) of chimera DTCs we consider a particular example of a one-dimensional array of $N=8$ spins. Here the coupling strength between spins is dependent on the distance they are apart with coupling strength $J_{{lm}}^{z} \equiv J_0/|{l}-{m}|^{\alpha}$ for the sites $l$ and $m$.
While the geometrical arrangement of spin is one dimensional, the parameter $\alpha$ determines the structure of the network. For example, if $\alpha=0$ the network is all-to-all connected and for $\alpha=\infty$, it has nearest-neighbors coupling only. It is important to note that our results are general and can be applied to other networks with more complex connectivities (see supplemental material ~\cite{SupplementalInfo}). Choosing an initial state $\ket{\Psi(0)}_z =\ket{1,1\cdots,1}_z $ which breaks the $\mathbb{Z}_2$-Ising symmetry, we can now explore, as illustrated in Fig.~\ref{Fig1a} (b), the effect of $\epsilon_A \ll1$ and $\epsilon_B \sim1$ on the systems magnetization. Here we set $T_1 =T_2 = T/2$ and determine the magnetization at $t_n = 2nT$.  Its dynamics is characterized by the ratio $\delta_x = \epsilon_A \pi(1+\overline{\langle \cos(\hat{\theta})  \rangle})/2J_0T$, where $\overline{\langle \cdots  \rangle}$ is the ensemble average. This indicates the balance between the effective transverse magnetic \blue{field} felt by A due to the coupling $\hat{H}_{\text{AB}}$, and the interaction strength $J_0$ .  It has a critical point at $\delta_x^c \sim 1$. We employ two different values of $\epsilon_A=0.1,0.03$ to investigate the dynamics above and below that critical point. The second term in $\hat{H}_{\epsilon_A,2T}^{\text{eff}}$ can be neglected as $J_0 \epsilon_A \ll 1$, however we need to consider disorder as another factor in the DTC's emergence (disorder is essential for the DTC to be stabilized under the imperfect rotations \cite{else2019discrete}). The short-time magnetization dynamics clearly shows two different behaviors in regions A and B as shown in Figs.~\ref{Fig1b}~(a, b) for the weak $(J_0T=0.072)$, strong $(J_0T=0.2)$ couplings respectively.  
In region A, the regional rotation breaks the discrete time translational symmetry yielding the DTC phase, while region B retains its ferromagnetic phase. It is also important to explore the long-time dynamics for different system sizes in the weak and strong coupling regimes  which we show in Figs.~\ref{Fig1b}~(c, d). It is clearly seen in this long-time regime that the DTC's melt~\cite{Machado2020,Zeng2017,Else2017}. Two coexistent phases of matter are still present.

\begin{figure}[t]
\centering
\includegraphics[width=0.45\textwidth]{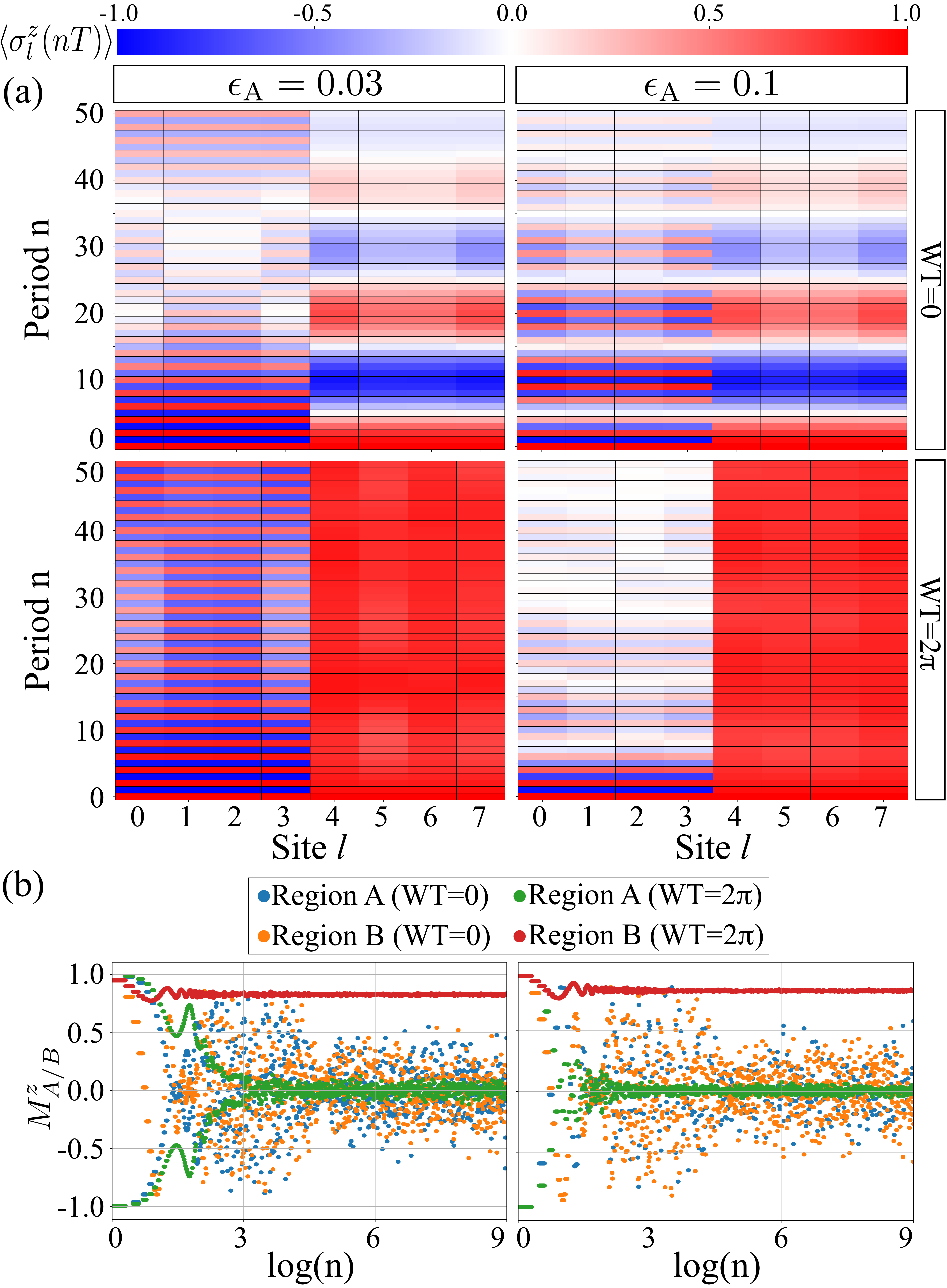}
\caption{ 
The stroboscopic evolution (a) of the ensemble averaged local magnetization $\langle\sigma_l^z(nT)\rangle$  over 100 realizations of disorder for a Chimera DTC in the weak coupling regime over short times. Here two specific values of $\epsilon_A$ are chosen $(\epsilon_A=0.03, 0.1)$ with disorder strengths $WT=0,2\pi$. The other parameters are the same as in Fig.~\ref{Fig1b} including $|\Psi(0)_z\rangle = |1,1\cdots,1\rangle$ as our initial state. In (b) we show the long-time dynamics of the regional magnetization $M^z_{A/B}$ for errors $\epsilon_\text{A}=0.03, 0.1$ (left, right panels), respectively. Here, the blue and green curves represent
 the regional magnetization $M^z_{A}$ of region A for disorder strengths $WT=0, 2\pi$ respectively, while the orange and red curves represent the regional magnetization $M^z_{B}$ in region B for $WT=0, 2\pi$.  The regional magnetization $M^z_{A}$ is periodic with period $2T$, whereas $M^z_{B}$ is constant under the effect of disorder.  This leads to the coexistence in space of the DTC and ferromagnetic phase.
} 
\label{Fig2}
\end{figure}

It is important to explore the effect of rotation errors and disorder on our Chimera DTC. Such effects are more prominent in the weak coupling regime.  In Fig.~\ref{Fig2} we plot the magnetization's dynamics for different values of $\epsilon_A$ and disorder strengths $W_l$. These are chosen randomly in the interval $[0,W]$, where $WT=0$ correspond to the no disorder case, while $WT=2\pi$ is strong disorder. We employ $100$ realizations in determining our ensemble average. In the regime  $\delta _x< 1$ ($\epsilon_A=0.03$) our Chimera DTC emerges. Increasing $W$ stabilized the DTC in region A. This is not unexpected as $\hat{H}_{\epsilon_A,2T}^{\text{eff}}$ can be seen as the Ising model perturbed by an effective transverse magnetic field $\hat{H}_{\text{AB}}$. The weaker the effective magnetic field (larger $W$), the more stable the DTC is.

In more detail, in the absence of disorder $WT=0$, region A feels a uniform transverse field with strength $\hbar \pi  \epsilon_A /2T$ and the value $\delta_x = \epsilon_A \pi/J_0T$ for $\epsilon_A = 0.03$ is effectively close to the critical point.  Due to the effect of the transverse magnetic field, the paramagnetic phase wins over the DTC phase as the time goes on, and the chimera DTC transforms into a new chimera phase of paramagnetic and ferromagnetic phases. With disorder $\langle \overline{\cos(\hat{\theta}) } \rangle \approx \langle \overline{\sin(\hat{\theta}) } \rangle \approx 0$ meaning the magnetic field is effectively suppressed.  In this case, $\delta_x = \epsilon_A \pi/2J_0T$ remains effectively small and the DTC phase of region A becomes stable.  These effects are shown in see Fig.~\ref{Fig2} (a). 
When $\delta_x > 1$ ($\epsilon_A =0.1$) the transverse magnetic field is dominant in the effective Hamiltonian $\hat{H}_{\epsilon,2T}^{\text{eff}}$ meaning the system exhibits many-body Rabi oscillations. In the absence of disorder, the local magnetization at each site of the region A oscillates with the same frequency. On the contrary, the disorder randomises the frequencies of the Rabi oscillations smoothing it out when the ensemble average is taken. Such behavior is illustrated in Fig.~\ref{Fig2} (a,b) for the short- and long-time dynamics in the weak coupling regime. As it is shown in Fig.~\ref{Fig2} ~(a), in the absence of disorder the spins in region B also oscillate with the same frequency. This means that the disorder stabilizes the magnetizations in region B at stroboscopic times.

\begin{figure*}[t]
\centering
\includegraphics[width=0.95\textwidth]{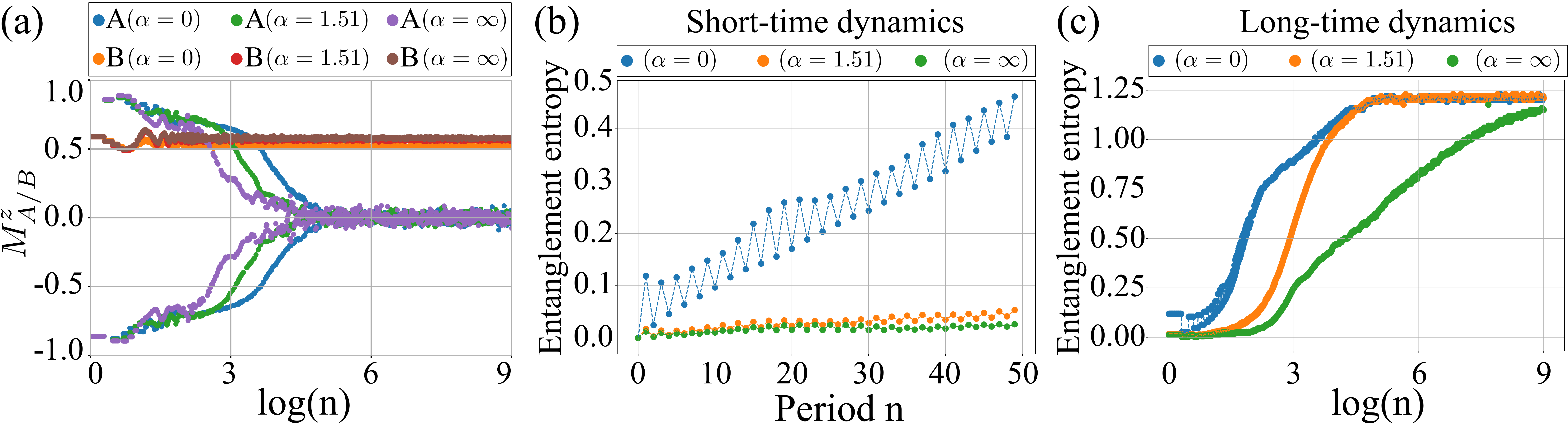}
\caption{
In (a) we plot the ensemble average of the regional magnetizations $M^z_{A/B}$ versus stroboscopic time for $\alpha=0,1.51,\infty$. We have used our typical strong coupling regime parameters: $J_0T = 0.2, \epsilon_A = 0.03, \epsilon_B = 0.9$, $gT = \pi$ and with disorder strength $WT = 2\pi$. Next (b,c) illustrate the short- and long-time entanglement $S_B(nT)$ dynamics. Our ensemble averaging utilizes 100 realizations of the disorder with the initial state $ |\Psi(0)\rangle_{z,\theta} = e^{ -i \sum_{l} \frac{ \theta }{2} \sigma_l^x } |\Psi(0)\rangle_{z}$ where $\theta=0.2\pi$.
}
\label{Fig4}
\end{figure*}

The above analysis has shown that the chimera phase can be observed from the magnetization dynamics of our quantum system and is stable for $\delta_x < 1$.  It is useful at this stage to focus our attention in the strong coupling regime (less sensitive to rotational errors) to explore the effect of long-range correlations and how they affect  the observed macroscopic behavior.  The archetypical quantum behavior is of course entanglement. Given our system remains pure state  throughout its evolution, we can evaluate the degree of entanglement~\cite{Horodecki2009,Eisert2010} between the two regions A and B using the von Neumann entropy $S_B(t) = - \Tr_B [\hat{\rho}_B (t) \ln {\hat{\rho}_B (t) }]$ where $\hat{\rho}_B (t) = \Tr_A [|\Psi(t)\rangle\langle \Psi(t)|] $ is the reduced density matrix of region B (other entanglement measures could be used if desired). In our exploration of entanglement in this Chimera DTC we need to consider both the effect of rotation errors $\epsilon_{A,B}$ and errors in the the initial state preparation. For the later case we will simply model our initial state as $ |\Psi(0)\rangle_{z,\theta} = e^{ -i \sum_{l} \frac{ \theta }{2} \sigma_l^x } |\Psi(0)\rangle_{z}$, where $\theta$ is the error. As $|\Psi(0)\rangle_{z,\theta \neq 0}$ are not eigenstates of $\hat{H}(t)$ we expect interesting dynamics to arise. In Fig.~\ref{Fig4}~(a) we plot the regional magnetization $M^z_{A/B}$ versus time for various $\alpha$ (interaction range) by considering $\epsilon_A=0.03$, $\epsilon_B=0.9$ and $\theta=0.2\pi$ with disorder strength $WT=2\pi$. We compare the magnetization dynamics for three different $\alpha$'s ($\alpha = 0,1.51,\infty$)  where $\alpha=1.51$ was chosen based on the recent experiments~\cite{zhang17}.  The $\alpha =0$ and $\alpha=\infty$ values correspond to all-to-all coupling and nearest-neighbor coupling respectively. Fig.~\ref{Fig4}~(a) shows that the Chimera DTC is robust against errors and two phases of matter (DTC and ferromagnetic) coexist for a long-time that depends on the interaction range $\alpha$. This shows that the Chimera DTC is more stable for all-to-all coupling $(\alpha=0)$ where it has the longest lifetime.

So far we have shown that the Chimera DTC is robust agains errors in both rotations and state preparation. The natural question that arises is how the degree of entanglement between regions A and B varies with long-range connectivity and the effect of rotation errors on it. In Fig.~\ref{Fig4}~(b)~and~(c) we plot the short- and long-time entanglement dynamics for three $\alpha$ values. For times where the DTC is stable [see Fig.~\ref{Fig4}~(b)] the degree of entanglement is small for short-range interactions $(\alpha=1.51,\infty)$ but increases as more long-range interactions are included $(\alpha=0)$. In the long-time limit, the DTC in region A melts and the entanglement reaches a steady state value that depends on the interaction range $\alpha$. 
The reason are local conserved quantities within the DTC phase that prevent long-range correlations between regions A and B. When the DTC in region A melts, quantum tunneling is possible between the different localized states resulting in an increase in entanglement between regions A and B. This is shown in Fig.~\ref{Fig4}~(c). Here it is important to compare our results with the entanglement entropy for generic thermal and manybody localized (MBL) states. If the system is thermalized, the average entanglement entropy is predicted to be $\overline{\langle S \rangle} \sim (N\ln(2)-1)/2\approx 2.3$ for $N=8$. On the contrary, in the MBL phase the predicted average entropy should be $\overline{\langle S \rangle} \sim \ln(2) \approx 0.69$. In our Chimera DTC, the interaction range $\alpha$ controls the degree of the correlations between regions A and B. For all the $\alpha$'s, the entanglement entropy eventually converges to an intermediate value $\overline{\langle S \rangle} \sim 1.23$. This means that even when the DTC in region A is melted, the entanglement entropy of our system lies between the values of a fully thermal and MBL states.

In  summary, we have shown how regional driving on quantum spin networks can manipulate the phases of matter associated with it. In particular we have demonstrated how chimera DTC's can emerge using that regional driving and are stable to imperfections. Of course, we are not restricted to two regions and can apply drives in multiple regions to generate even more complex Chimera-like phases. We could for instance create Chimera phases composed of distinct DTC's surrounded by ferromagnetic or even ergodic domains.  The separability of these distinct phase regions is not required for the chimera DTC's to appear. Entanglement can be present at the stroboscopic times  $2nT$.  It is also interesting that in the chimera DTCs the amount of entanglement is rather suppressed despite the broad interactions across the spin network.  This suggests that the chimera DTCs may be used to control subsystems of a spin network.  Our results are experimentally realizable with the quantum technologies available today in various platforms including superconducting circuits, trapped ions and cold atoms.

We thank M. P. Estarellas, and T. Osada for valuable discussions. V. M. B. acknowledges fruitful discussions with E. Sch\"oll. This work was supported in part from the Japanese  MEXT  Quantum  Leap  Flagship  Program  (MEXT  Q-LEAP)  Grant  No.JPMXS0118069605, the MEXT KAKENHI Grant-in-Aid for Scientific Research on Innovative Areas Science of hybrid quantum systems Grant No.15H05870 and the JSPS KAKENHI Grant No. 19H00662.

\newpage
\clearpage
\onecolumngrid

\section{Supplementary materials for ``Chimera Time-Crystalline order in quantum spin networks"} 

\subsection*{
A roadmap for the supplemental information
\label{Sec0}
}

The purpose of this supplemental information is to provide additional details on the analytical and numerical calculations in the limiting case $\epsilon_{\text{B}}=1$. With this aim, we decided to divide the supplemental material into four main parts. In section~I we show in detail how to analytically obtain the effective Hamiltonian $\hat{H}_{\epsilon_A,2T}^\text{eff}$ for non-zero error $\epsilon_\text{A}$ in the case of our Chimera DTC.  In section~II we focus on the strong coupling regime and discuss the long-time dynamics in the limiting case $\epsilon_{\text{B}}=1$. In this case, there are several local symmetries restricting the dynamics of the system. Thus, the region B is always in the ferromagnetic phase and it is protected by conservation laws against the effect of the drive. In section~III, we turn our attention to the weak coupling regime and the short-time dynamics. Again, we focus on the limiting case $\epsilon_{\text{B}}=1$ and show the effect of the rotation error  $\epsilon_{\text{A}}$ on region A. In section~IV, we discuss the dynamics of regional magnetization for small errors $\epsilon_A \leq \epsilon_B \ll 1$. This regime is interesting as the errors are small but still different for the regions A and B. In the final section~V, we provide additional numerical evidence of Chimera DTC in a network with a nontrivial connectivity: A ladder-like quantum spin network. This example is instructive, as it opens the possibility to study the dynamics of the Chimera DTC for different bipartitions A and B.

\section{
I. An effective Hamiltonian for a non-zero value of $\epsilon_\text{A}>0$ and $\epsilon_\text{B}=1$
\label{SecI}
}

In this section, we analytically derive the effective Hamiltonian $\hat{H}_{\epsilon_A,2T}^\text{eff}$ for $2T$ for a non-zero error $\epsilon_\text{A}$. In the calculation, we assume that $\epsilon_\text{A}$ is the small and set $\epsilon_\text{B} = 1.0$. Our derivation is valid for any dimension $d$ and connectivity of the quantum spin network. To find the effective Hamiltonian, we start from the square of the Floquet operator. By definition, in our model, it can be written as

\begin{equation}
\begin{split}
\hat{\mathcal{F}}^2 &= \exp\left(-\frac{i}{\hbar}\hat{H}_{\epsilon_A,2T}^\text{eff} 2T\right) = \exp\left(-\frac{i}{\hbar}\hat{H}_2 T_2\right)  \exp\left(-\frac{i}{\hbar}\hat{H}_1 T_1\right)  \exp\left(-\frac{i}{\hbar}\hat{H}_2 T_2\right) \exp\left(-\frac{i}{\hbar}\hat{H}_1 T_1\right) ,
\end{split}
\label{eq:Floquet2T}
\end{equation}
where $\hat{H}_1$ and $\hat{H}_2$ are defined in the main text. For convenience, we decompose $H_2$ as follows,
\begin{equation}
\begin{split}
	\hat{H}_2 &= \hbar \sum\limits_{lm}J_{lm}^{z} \sigma_{l}^{z} \sigma_{m}^{z} +  \hbar\sum\limits_{l}W_{l}^{z} \sigma_{l}^{z} \\
	&=\hbar \sum\limits_{lm \in \text{A} } J_{lm}^z \sigma_{l}^z \sigma_{m}^z
+ \hbar\sum\limits_{lm \in \text{B} } J_{lm}^z \sigma_{l}^z \sigma_{m}^z
+\hbar \sum\limits_{l \in \text{A}, m \in \text{B} } J_{lm}^z \sigma_{l}^z \sigma_{m}^z 
+ \hbar \sum\limits_{ l\in \text{A}} W_l^z \sigma_l^z 
+ \hbar \sum\limits_{  l \in \text{B}} W_l^z \sigma_l^z 
\ .
\end{split} 
\end{equation}
Using the properties of the spin rotation we rewrite the operator $\hat{O}=\exp\left(-\frac{i}{\hbar}\hat{H}_1 T_1\right)  \exp\left(-\frac{i}{\hbar}\hat{H}_2 T_2\right) \exp\left(-\frac{i}{\hbar}\hat{H}_1 T_1\right)$, as follows
\begin{equation}
\begin{split}
\hat{O}  & = \exp \left[-\frac{i\pi (1-\epsilon_{\text{A}}  )}{2}\sum\limits_{l\in \text{A}} \sigma_{l}^x\right]   \exp \left[-i \left(  \sum\limits_{lm}J_{lm}^{z} \sigma_{l}^{z} \sigma_{m}^{z} +  \sum\limits_{l}W_{l}^{z} \sigma_{l}^{z} \right) T_2\right]  \exp \left[-\frac{i\pi (1-\epsilon_{\text{A}} )}{2}\sum\limits_{l\in \text{A}} \sigma_{l}^x \right]  \\
&= \hat{V}_{\epsilon_A}\exp \left[-i \left(  \sum\limits_{lm \in \text{A} } J_{lm}^z \sigma_{l}^z \sigma_{m}^z
+ \sum\limits_{lm \in \text{B} } J_{lm}^z \sigma_{l}^z \sigma_{m}^z
-  \sum\limits_{l \in \text{A}, m \in \text{B} } J_{lm}^z \sigma_{l}^z \sigma_{m}^z 
- \sum\limits_{l\in \text{A}} W_{l}^z \sigma_{l}^z 
+ \sum\limits_{l\in \text{B}} W_{l}^z \sigma_{l}^z 
\right)T_2\right]
\hat{V}_{\epsilon_A} \\
&= \hat{V}_{\epsilon_A}
\exp \left[-i \left(  \sum\limits_{lm \in \text{A} } J_{lm}^z \sigma_{l}^z \sigma_{m}^z
+\sum\limits_{lm \in \text{B} } J_{lm}^z \sigma_{l}^z \sigma_{m}^z
-  \sum\limits_{{l} \in \text{A}} ( W_{l}^z + \sum\limits_{{m} \in \text{B}} J_{lm}^z \sigma_{m}^z) \sigma_{l}^z  
+ \sum\limits_{{l}\in \text{B}} W_{l}^z \sigma_{l}^z 
\right)T_2 \right]
\hat{V}_{\epsilon_A} \\
&= \hat{V}_{\epsilon_A}
\exp \left[i \left(   \sum\limits_{{l} \in \text{A}} ( W_{l}^z + \sum\limits_{m \in \text{B}} J_{lm}^z \sigma_m^z) \sigma_l^z  
 \right)T_2  \right]
\exp \left[-i \left(  \sum\limits_{lm \in \text{A} } J_{lm}^z \sigma_{l}^z \sigma_{m}^z  + \hat{H}_B
\right)T_2 \right]
\hat{V}_{\epsilon_A},
\end{split}
\label{eq:rotation}
\end{equation}
where $\hat{V}_{\epsilon_A}=\exp\left(\frac{i\epsilon_\text{A} \pi }{2} \sum\limits_{{l} \in \text{A}} \sigma_{l}^x \right)$ and the Hamiltonian $\hat{H}_B = \sum\limits_{lm \in \text{B} } J_{lm}^z \sigma_{l}^z \sigma_{m}^z
+ \sum\limits_{{l}\in \text{B}} W_{l}^z \sigma_{l}^z $ contains operators of the region B.
Let us define $ \hat{\theta}_{l} = (2 W_{l}^z T_2+ 2T_2\sum_{{m} \in \text{B}} J_{lm}^z \sigma_{m}^z) $, which can be used to define a rotation acting on the operator $\sigma_{l}^z$, 
\begin{equation}
\begin{split}
\exp \left(-i  \sum\limits_{{l} \in \text{A}}\frac{\hat{\theta}_{l}}{2} \sigma_{l}^z \right) 
\hat{V}_{\epsilon_A}
\exp \left(i  \sum\limits_{{l} \in \text{A}} \frac{\hat{\theta}_{l}}{2}  \sigma_l^z \right)   
&= \exp \left[\frac{i\epsilon_\text{A} \pi }{2}  \sum\limits_{{l} \in \text{A}} \left( \cos { \hat{\theta}_{l} } \sigma_{l}^x + \sin{ \hat{\theta}_{l} } \sigma_{l}^y \right)  \right]
\ .
\end{split} 
\label{eq:alicerotation}
\end{equation}
Thus, 
\begin{equation}
\begin{split}
\hat{\mathcal{F}}^2 & = 
\exp \left[-i \left(  \sum\limits_{lm \in \text{A} } J_{lm}^z \sigma_{l}^z \sigma_{m}^z \right)T_2 \right]
\exp \left(-i  \sum\limits_{{l} \in \text{A}}\frac{\hat{\theta}_{l}}{2} \sigma_{l}^z \right) 
\hat{V}_{\epsilon_A}
\exp \left(i  \sum\limits_{{l} \in \text{A}} \frac{\hat{\theta}_{l}}{2}  \sigma_l^z \right) \exp \left[-i \left(  \sum\limits_{{lm} \in \text{A} } J_{{lm}}^z \sigma_{l}^z \sigma_{m}^z  +  2 \hat{H}_B
\right)T_2 \right]
\hat{V}_{\epsilon_A}
\\
&= \exp \left[-i \left(  \sum\limits_{lm \in \text{A} } J_{lm}^z \sigma_{l}^z \sigma_{m}^z \right)T_2 \right]\exp \left[\frac{i\epsilon_\text{A} \pi }{2}  \sum\limits_{{l} \in \text{A}} \left( \cos { \hat{\theta}_{l} } \sigma_{l}^x + \sin{ \hat{\theta}_{l} } \sigma_{l}^y \right)  \right]
\exp \left[-i \left(  \sum\limits_{lm \in \text{A} } J_{lm}^z \sigma_{l}^z \sigma_{m}^z \right)T_2 \right]\exp \left(-i 2 \hat{H}_B T_2 \right)
\hat{V}_{\epsilon_A}
\ .
\end{split}
\end{equation}
When the parameter $\lambda$ and $\mu$ are small, the Baker-Campbell-Hausdorff formula~\cite{scharf1988campbell} is approximately given by
\begin{equation}
\exp\left(\lambda\hat{X}\right)\exp\left( \mu\hat{Y}\right) \approx \exp\left(\lambda\hat{X}+\mu \hat{Y} + \frac{1}{2} \lambda\mu  [\hat{X},\hat{Y}]\right),
\end{equation}
where $\hat{X}$ and $\hat{Y}$ are matrices. By using this equation, one can write 
\begin{equation}
\begin{split}
\exp\left(\lambda\hat{X}\right)\exp\left(\mu \hat{Y}\right) \exp\left(\lambda\hat{X}\right)&= \exp\left(\lambda\hat{X}\right)\exp\left(\frac{\mu}{2} \hat{Y}\right) \exp\left(\frac{\mu}{2} \hat{Y}\right)\exp\left(\hat{X}\right) \\
& \approx  \exp\left(\lambda\hat{X} + \frac{\mu}{2} \hat{Y} + \frac{\lambda\mu}{4} [\hat{X},\hat{Y}]\right) \exp\left(\lambda\hat{X} + \frac{\mu}{2} \hat{Y} - \frac{\lambda\mu}{4} [\hat{X},\hat{Y}]\right) \\
& \approx  \exp\left(2\lambda\hat{X} + \mu\hat{Y} \right).
\end{split}
\end{equation}
We here assume that both the couplings $J_{{lm}}^z$ as well as the error $\epsilon_A$ are small. Thus, we can apply the formula discussed above to obtain 

\begin{equation}
\begin{split}
&\exp \left[-i \left(  \sum\limits_{lm \in \text{A} } J_{lm}^z \sigma_{l}^z \sigma_{m}^z \right)T_2 \right]\exp \left[\frac{i\epsilon_\text{A} \pi }{2}  \sum\limits_{{l} \in \text{A}} \left( \cos { \hat{\theta}_{l} } \sigma_{l}^x + \sin{ \hat{\theta}_{l} } \sigma_{l}^y \right)  \right]
\exp \left[-i \left(  \sum\limits_{lm \in \text{A} } J_{lm}^z \sigma_{l}^z \sigma_{m}^z \right)T_2 \right] \\
&= \exp \left[-i \left( 2 \sum\limits_{lm \in \text{A} } J_{lm}^z \sigma_{l}^z \sigma_{m}^z \right)T_2 +\frac{i\epsilon_\text{A} \pi }{2}  \sum\limits_{{l} \in \text{A}} \left( \cos { \hat{\theta}_{l} } \sigma_{l}^x + \sin{ \hat{\theta}_{l} } \sigma_{l}^y \right)\right]
\ .
\label{eq:approx}
\end{split}
\end{equation}
By combining Eqs \eqref{eq:rotation}, \eqref{eq:alicerotation} and \eqref{eq:approx}, finally, we get 

\begin{equation}
\begin{split}
	\hat{H}_{\epsilon_A,2T}^{\text{eff}} &\approx  \frac{\hbar T_2}{T} \sum_{l,m\in A} J_{lm}^z \sigma_l^z \sigma_m^z - \frac{\hbar \pi \epsilon_A   T_2} {2T}\sum_{\substack{l,m \in A \\  l \neq m }} J_{lm}^z \sigma_l^z \sigma_m^y   - \frac{\hbar \pi  \epsilon_A }{4T} \sum_{l \in A}  \Big[
	\Big(  \cos{\big(2W_l T_2 + 2\sum_{m\in B} J_{lm}^z T_2 \sigma_m^z \big)} + 1 \Big) \sigma_l^x \\
	 & \quad+
	 \sin{\big(2W_l T_2 + 2\sum_{m\in B} J_{lm}^z T_2 \sigma_m^z \big)}  \sigma_l^y \Big]  + \frac{\hbar T_2}{T} \sum_{l,m\in B}J_{lm}^z \sigma_l^z \sigma_m^z + \frac{\hbar T_2}{T} \sum_{l \in B} W_l \sigma_l^z.
	 \label{eq:FullEffHam}
\end{split}
\end{equation}
From this Hamiltonian, we see that the rotation error $\epsilon_\text{A}$ produces an interaction between regions A and B, and the strength of this interaction is proportional to it.  If $\epsilon_\text{A}=0$, this Hamiltonian reduces to  \begin{equation}
\hat{H}_{\epsilon_A=0,2T}^{\text{eff}} =
 \frac{\hbar T_2}{T} \sum_{\boldsymbol{l,m} \in \text{A}} J_{lm}^z \sigma_{l}^z \sigma_{m}^z 
 + \frac{\hbar T_2}{T} \sum_{\boldsymbol{l,m}\in \text{B}}  J_{lm}^z \sigma_{l}^z \sigma_{m}^z  
 +  \frac{\hbar T_2}{T}\sum_{l\in \text{B}} W_{l}^z \sigma^z_{l},
\end{equation} 
which is the same as the exact one obtained from Eq. \eqref{eq:Floquet2T} analytically.  Obviously, regions A and B are completely decoupled. 

\subsection{A. The Ensemble average of entanglement entropy }

In the numerical calculation of the entanglement entropy with the rotation error, we exploit the assemble entanglement entropy to take with the randomness in the disorder statistically into account. We define the ensemble average entropy $\overline{\langle  S_B (t) \rangle}$ at time $t$ as
\begin{equation}
	\overline{\langle  S_B (t) \rangle}= \frac{1}{L}\sum_{r}^{L} S_B^{(r)}(t),
	\label{eq:ensembleentropy}
\end{equation}
where $L$ is the total number of realizations of disorder and $S_B^{(r)}(t)$ is the von Neumann entropy for a single realization $r$ at time $t$. 

\subsection{B. Entanglement dynamics and effective dephasing }

As we explain in detail in Eq.~$(2)$ of the main text, the effective Hamiltonian has a structure  $\hat{H}_{\epsilon_A,2T}^{\text{eff}}=\hat{H}_{\text{A}}+\hat{H}_{\text{B}}+\hat{H}_{\text{AB}}$, where $\hat{H}_{\text{A}}$ and  $\hat{H}_{\text{B}}$ are effective Hamiltonians for A and B and $\hat{H}_{\text{AB}}$ describe their coupling.  
The  coupling  is determined by operator $ \hat{\theta}_{l}=2W_{l} T_2 + 2T_2\sum_{m\in B} J_{lm}^z \sigma_{m}^z $with $l \in \text{A}$, because it contains operators of the subsystem B. The system B is governed by the Hamiltonian

\begin{equation}
         \label{eq:EffHamB}
\hat{H}_{\text{B}}=\frac{\hbar}{2} \sum_{\boldsymbol{l,m} \in B}J_{\boldsymbol{l,m}}^z \sigma_{l }^z \sigma_{m }^z + \frac{\hbar}{2} \sum_{l \in B} W_{l } \sigma_{\boldsymbol{l,m} }^z
\ .
\end{equation}

Remarkably, this operator is a conserved quantity of the stroboscopic dynamics and $[\hat{H}_{\text{B}},\hat{H}_{\epsilon_A,2T}^{\text{eff}}]=0$. This also has important consequences on the entanglement dynamics. When we prepare the system in an initial state $|\Psi(0)\rangle_z = |1,1\cdots 1\rangle$, we find that

\begin{align}
	\hat{H}_{\epsilon_A,2T}^{\text{eff}}|\Psi(0)\rangle_z &=  \left(\frac{\hbar T_2}{T} \sum_{\boldsymbol{l,m}\in A} J_{lm}^z  - \frac{\hbar \pi \epsilon_A } {4}\sum_{\substack{\boldsymbol{l,m} \in A }} J_{\boldsymbol{l,m}}^z  \sigma_{m}^y\right)|\Psi(0)\rangle_z
	\nonumber\\&
	  +\left( \frac{\hbar T_2}{T} \sum_{\boldsymbol{l,m} \in B}J_{\boldsymbol{l,m} }^z  + \frac{\hbar }{2} \sum_{l \in B} W_{l } \right)|\Psi(0)\rangle_z
	\nonumber\\&  
	- \left(\frac{\hbar \pi  \epsilon_A T_2 }{4T} \sum_{l  \in A}  \Big[
	\Big(  \cos{\langle\hat{\theta}_{l }\rangle} + 1 \Big) \sigma_{l }^x +
	 \sin{\langle\hat{\theta}_{l }\rangle}  \sigma_{l }^y \Big] \right)|\Psi(0)\rangle_z
	  \label{eq:ActioneffHamil2T}
	  \ .
\end{align}

\begin{figure}[h]
\centering
\includegraphics[width=0.6\textwidth]{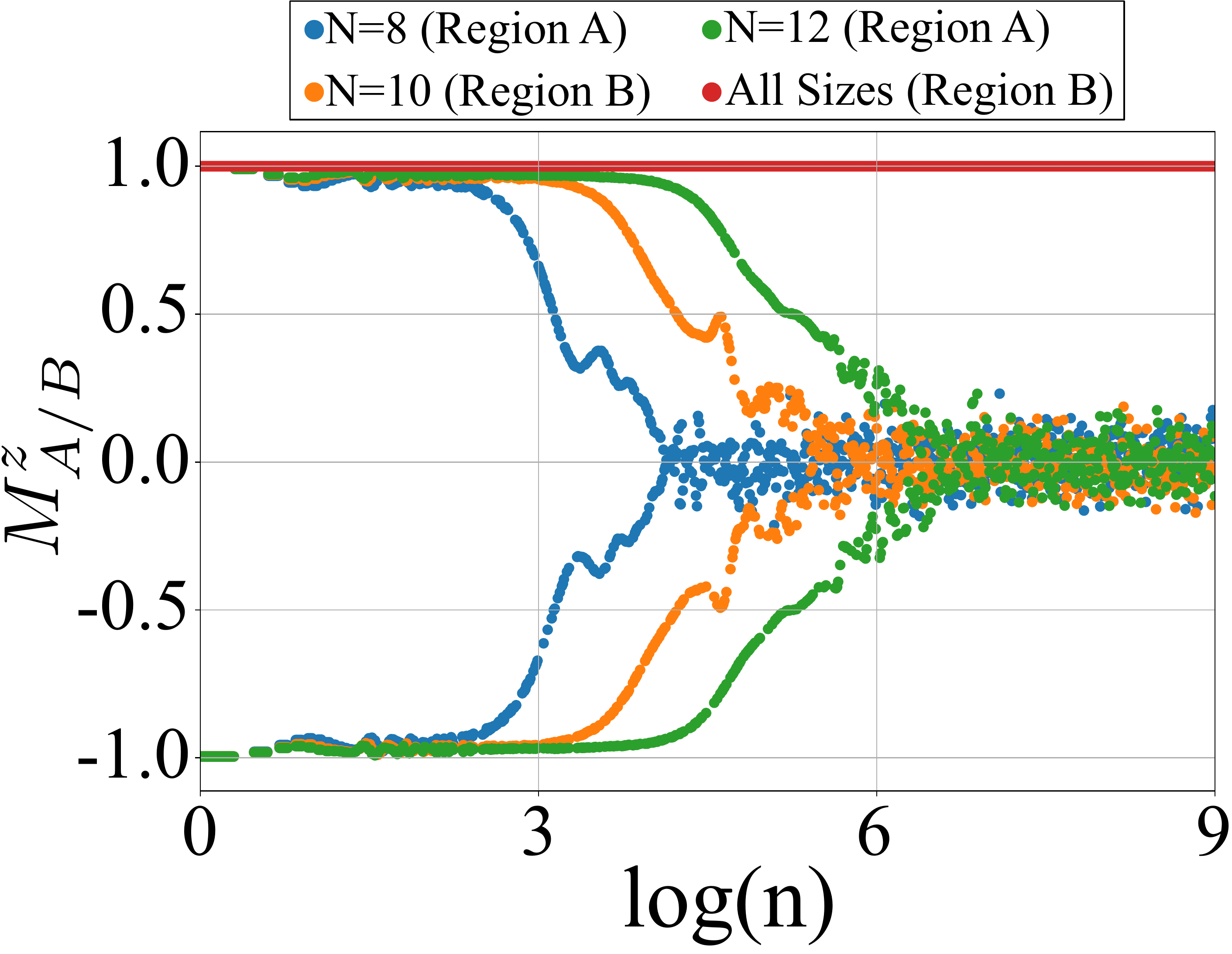}
\caption{Long-time dynamics of the ensemble averaged regional magnetization $M^z_{A/B}$ for rotation errors $\epsilon_A=0.03$ and $\epsilon_B = 1$. We calculate the ensemble average by using 100 realizations of disorder. We set $gT= \pi$, $J_0T = 0.2$, $\alpha = 1.51$ and $WT=2\pi$. 
}
\label{Fig1S}
\end{figure}

\begin{figure}[h]
\centering
\includegraphics[width=0.50\textwidth]{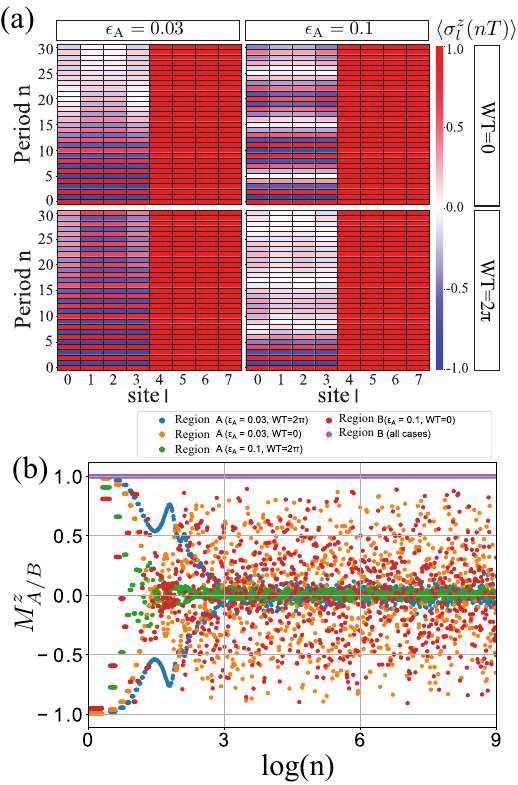}
\caption{ 
The stroboscopic evolution (a) of the ensemble averaged local magnetization $\langle\sigma_l^z(nT)\rangle$  over 100 realizations of disorder for a Chimera DTC in the weak coupling regime $J_0T=0.072$ over short times and for $\epsilon_B=1$. Here two specific values of $\epsilon_A$ are chosen $(\epsilon_A=0.03, 0.1)$ with disorder strengths $WT=0,2\pi$. We prepare $|\Psi(0)_z\rangle = |1,1\cdots,1\rangle$ as our initial state. In (b) we show the long-time dynamics of the regional magnetization $M^z_{A/B}$ for different errors $\epsilon_\text{A}=0.03, 0.1$ and disorder strengths $WT=0, 2\pi$.  In the short-time limit, the regional magnetization $M^z_{A}$ is periodic with period $2T$, whereas $M^z_{B}$ is constant.  This leads to the coexistence in space of the DTC and ferromagnetic phase. We set $gT= \pi$, and $\alpha = 1.51 $.
}
\label{Fig2S}
\end{figure}
Note that we have used  that $\hat{H}_{\text{B}}|\Psi(0)\rangle_z=E_{\text{B}}|\Psi(0)\rangle_z$, where $E_{\text{B}}= \frac{\hbar T_2}{T} \sum_{\boldsymbol{l,m} \in B}J_{\boldsymbol{l,m} }^z  + \frac{\hbar }{2} \sum_{l \in B} W_{l } $. We also have defined the mean rotation angle $\langle\hat{\theta}_{l }\rangle=2W_{l} T_2 + 2T_2\sum_{m\in B} J_{lm}^z  $. As one can see from this discussion, the Hamiltonian cannot create entanglement between the regions A and B at stroboscopic times. Next, let us discuss what happen when we prepare the initial state 
$|\Psi(0)\rangle_x = |+  +\cdots+ \rangle $, which is fully polarized along $x$-axis, where $|+\rangle = \frac{1}{\sqrt{2}} (|0 \rangle + |1 \rangle)$ is an eigenstate of $\sigma_l^x$ at the $l$-th site. The state $\Psi(0)\rangle_x$ can be written as a linear combination of eigenstates of $\sigma_l^z$. This resembles dephasing dynamics, because the system A acts as an environment and affect the phase information of quantum superpositions in the system B.

\section{
II. Long-time dynamics in the strong coupling regime $J_0T=0.2$ for different system sizes $N$ and fixed rotation error $\epsilon_{\text{B}}=1$
\label{SecII}
}

One of the most important aspects of discrete time crystals (DTCs) is symmetry breaking. In fact, a DTC is a phase of matter that breaks discrete translational symmetry in time. however, in numerical studies we usually restrict ourselves to finite systems. For the short-time dynamics this has not much effect because if we prepare the system in a state that breaks the symmetry of the Hamiltonian, the system will take a long time to tunnel to other configurations in the Hilbert space. Nevertheless, the finite size of the system has implications in the long-time dynamics of a DTC.

Keeping this in mind, in this section we investigate the dynamics of a Chimera DTC for different system sizes in the strong coupling regime $J_0T=0.2$.
We consider an important limiting case of our work by fixing $\epsilon_{\text{B}}=1$ and investigating the effect of a small error $\epsilon_{\text{A}}=0.03$. For these error values, the regions A and B are in the DTC and ferromagnetic phases, respectively. Due to the strong coupling, the DTC is stable for several periods of the drive in the prethermal regime~\cite{Machado2020, Zeng2017, Else2017}. However, as we previously discussed, due to finite size effects the system can tunnel to other configurations which leads to the melting of the DTC in region A. For a rotation error $\epsilon_{\text{B}}=1$ in region B, both the local magnetizations $\sigma^z_l$ for sites $l$ in B and the effective Hamiltonian Eq.~\eqref{eq:EffHamB} are conserved quantities. These conservation laws strongly influence the dynamics because they restrict the available configurations of the Hilbert space that the system can explore over time.

In Fig.~\ref{Fig1S} we plot the long time dynamics of the regional magnetization $M^z_{A/B}=2/N\sum_{l\in ={A/B}}\langle\sigma^z_l (nT)\rangle$ for our Chimera DTC as a function of $\log(n)$, where $n$ is the number of periods. From this figure we clearly see that there are two different coexisting phases at all time scales. For short times, the DTC coexist with the ferromagnetic phase and for long times, the DTC in region A melts and still coexist with the ferromagnetic phase in region B, which is protected by conservation laws. Remarkably, our numerical results confirm that the Chimera DTC is stable during a time scale that depend on the system size. The bigger the system, the longer is the lifetime of the DTC in region A.

\section{
III. Short-time dynamics in the weak coupling regime $J_0T=0.072$ for different system sizes $N$ and fixed rotation error $\epsilon_{\text{B}}=1$
\label{SecIII}
}

Our aim in this section is to discuss the dynamics of the system for short times and a fixed system size $N=8$. We focus here on the weak coupling regime and explore the error $\epsilon_{\text{A}}$ on the dynamics of the DTC in region A. Due to our choice of the error $\epsilon_{\text{B}}=1$, the region B is always in the ferromagnetic phase and its not affected by the drive. On the contrary, due to the weak coupling $J_0T=0.072$ considered here, the rotation error $\epsilon_{\text{A}}$ plays an important role because it can melt the DTC in region A. On the other hand, in this regime we can appreciate the effect of disorder on the stability of our Chimera DTC.

As we discussed in the main text, the region A feels an effective transverse field due to the coupling to region B that is proportional to the rotation error $\epsilon_{\text{A}}$. When this is larger than the coupling between the spins, the DTC rapidly melts. On the contrary, for error $\epsilon_{\text{A}}$ smaller than the interaction, the DTC is stable. Interestingly, in either case, there is a Chimera state where two phases of matter coexist in different regions of space. In Fig.~\ref{Fig2S}~(a) we show the short-time dynamics of the local magnetization $\langle\sigma^z_l (nT)$. Correspondingly in Fig.~\ref{Fig2S}~(b) we plot the dynamics of the regional magnetization $M^z_{A/B}=2/N\sum_{l\in ={A/B}}\langle\sigma^z_l (nT)\rangle$ for our Chimera DTC as a function of $\log(n)$. There we depict the dynamics for different values of the error $\epsilon_{\text{A}}$ and different disorder strengths.

\section{IV. 
Magnetization dynamics in the periodically driven system for intermediate error region $\epsilon_A \leq \epsilon_B \ll 1$ 
\label{SecIV}
} 
\begin{figure*}[h]
\centering
\includegraphics[width=0.95\textwidth]{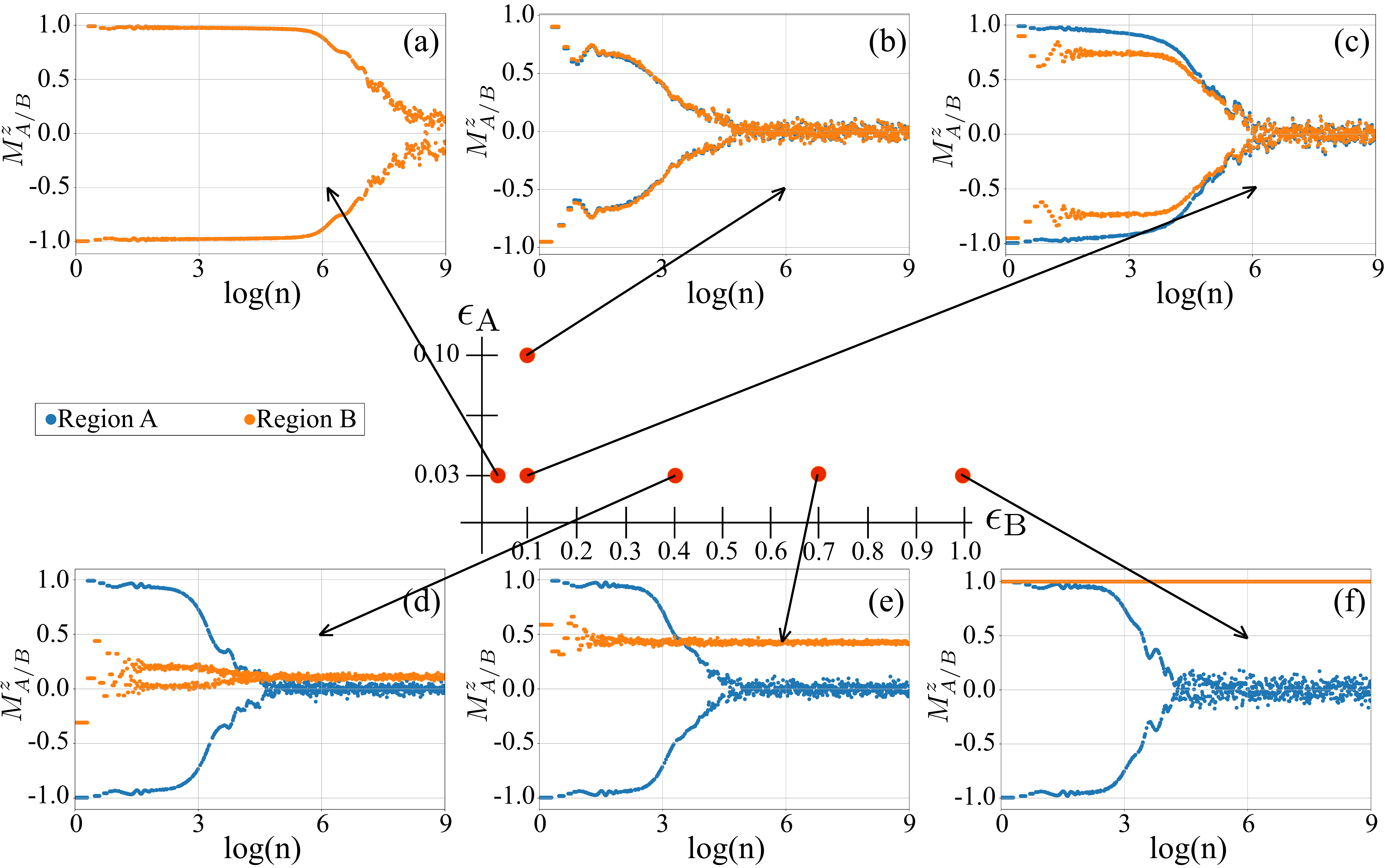}
\caption{ 
Dynamics of ensemble averaged regional magnetization $M^z_{A/B}$ for rotation errors $\epsilon_A \leq \epsilon_B \leq 1$:  We employ a one dimensional spin chain with $N=8$ sites in which sites from the $0$-th to $3$-rd are region A, and other sites are region B. To deal with randomness of the disorder, we calculate the ensemble average of the magnetization with 100 realizations. In panels (a) and (c)-(f), we fix the value of the error in region A  to a small value $\epsilon_A=0.03$ , and investigate the effect of the error $\epsilon_B$ on the dynamics. As long as $\epsilon_A$ is small enough, the whole system behaves like a DTC. For example, in (c) we depict the dynamics for non-uniform errors $\epsilon_A=0.03$ and $\epsilon_B=0.1$. For these error values the uncoupled regions A and B are in the DTC and ergodic phases, respectively. The coupling between the regions stabilizes system which behaves as a DTC. Contrary to this, when $\epsilon_B$ is significantly large, the magnetization dynamics exhibits chimera nature and two different phases of matter coexist in space. We set $gT= \pi$, $J_0T = 0.2$, $\alpha = 1.51$ and $WT=2\pi$. 
}
\label{Fig3S}
\end{figure*}

In the main text, we have described the dynamics governed by Eq. (1) for $\epsilon_A \ll 1$ and  $\epsilon_B \sim 1$. We also have shown the emergence a chimera DTC in which DTC and ferromagnetic phases coexist. This is possible in the regime in $\delta_x \ll 1$, where $\delta_x$ is a characteristic parameter for the chimera DTC. In this section, we show the dynamics for intermediate error regime ($\epsilon_A \leq \epsilon_B \ll 1$). To proceed with a simple analysis of this situation, we consider a one-dimensional disordered spin chain. We focus here on the strong coupling regime $J_0T = 0.2$, and set $\alpha = 1.51$ and $WT=2\pi$ for the interaction range and the disorder strength, respectively.  The results of the ensemble averaged (over 100 realizations of disorder) regional magnetization $M^z_{A/B}$ dynamics for different rotational errors are shown in Fig.~\ref{Fig3S}.

To understand the dynamics shown in Fig.~\ref{Fig3S} in a simple fashion, we analytically calculate  the effective Hamiltonian for two periods of the drive in the limit where two rotational errors are sufficiently small ($\epsilon_A,\epsilon_B \ll1$), as follows, 
\begin{align}
\hat{H}_{\epsilon_A \leq \epsilon_B\ll1,2T}^{\text{eff}} &\approx \frac{\hbar T_2}{T} \sum_{l,m} J^z_{lm} \sigma_l^z\sigma_m^z- \frac{\hbar\pi \epsilon_A}{4T}  \sum_{l\in A} \left[ \left( 1 + \cos{\left(2W_lT_2\right)} \right)\sigma_l^x  + \sin{(2W_l T_2)}\sigma_l^y \right] \nonumber\\&
- \frac{\hbar\pi \epsilon_B}{4T}  \sum_{l\in B} \left[ \left( 1 + \cos{\left(2W_lT_2\right)} \right)\sigma_l^x  + \sin{(2W_l T_2)}\sigma_l^y \right]
\end{align} 
It indicates that the dynamics in each regions can be characterized by parameters $\delta_x^{A/B}  \sim \pi\epsilon_{A/B}/{4J_0T_2}$ which measure the balance between the manybody interaction and the effective magnetic field acting on regions A and B. These parameters have a critical point $ \delta^{A/B}_{x,c}\sim 1$, and the dynamics before and after the critical point should be dramatically different. Here, let us apply these parameters to the simplest case in which the rotational errors are equal ($\epsilon_A = \epsilon_B$). When the rotational errors are relatively small $\epsilon_A = \epsilon_B\ll 1$, both regions shows the DTC that are relatively stable even at long periods, as shown in Fig.~\ref{Fig3S} (a). Contrary to this, if the parameter close to the critical point, or rotational errors are comparatively large, the DTC is melted over time, as shown in Fig.~\ref{Fig3S} (b). These results indicates that the analysis of dynamics by the parameter $\delta_x^{A/B}$ works well, when the rotational error is uniform.

Now, let us now investigate the dynamics when rotational errors not uniform ($\epsilon_A$ and $\epsilon_B$ take different values). Here, we fix $\epsilon_A$ that is a sufficiently small, and we discretely vary $\epsilon_B$. First, we consider $\delta_x^A < 1$ and $\delta_x^B\sim 1$ in which regions A and B are far and close to the critical point, respectively. We show the result in Fig.~\ref{Fig3S} (c). Surprisingly, by comparing Fig.~\ref{Fig3S} (b) and (c), we see that the stable DTC in region A prevents the melting of the DTC in region B in despite of the parameter $\delta_x^B\sim 1$ in region B being close to the critical point.  This indicates that there is a proximity effect due to the coupling between regions A to region B, and the interaction induces a DTC in region B. Thus, as long as $\delta_x^A \ll 1$, the whole system becomes a DTC even though $\delta_x^A < \delta_x^B\sim 1$.

Next, we consider the case in which $\epsilon_B$ is significantly large, or $\delta_x^B$ is much larger than the critical point.  
Fig.~\ref{Fig3S} (d) shows the results of the magnetization dynamics for $\epsilon_B = 0.4$. Now, the dynamics in region B shows a MBL paramagnetic-like dynamics due to the disorder, and the region A is still the DTC. This  means that in this regime ($0.1 \ll  \epsilon_B < 0.7$), there is a new chimera state in which the DTC and MBL paramagnetic phases coexist, which is different from the chimera state explained in the main text. 
Finally, when the rotational error $\epsilon_B$ is close to one, the dynamics in region B shows MBL ferromagnetic phase, as shown in Fig.~\ref{Fig3S} (e) and (f). In this regime, the system is a chimera system in which DTC and MBL ferromagnetic phases coexist, as we discussed in the main text. 

\begin{figure*}[h]
\centering
\includegraphics[width=0.60\textwidth]{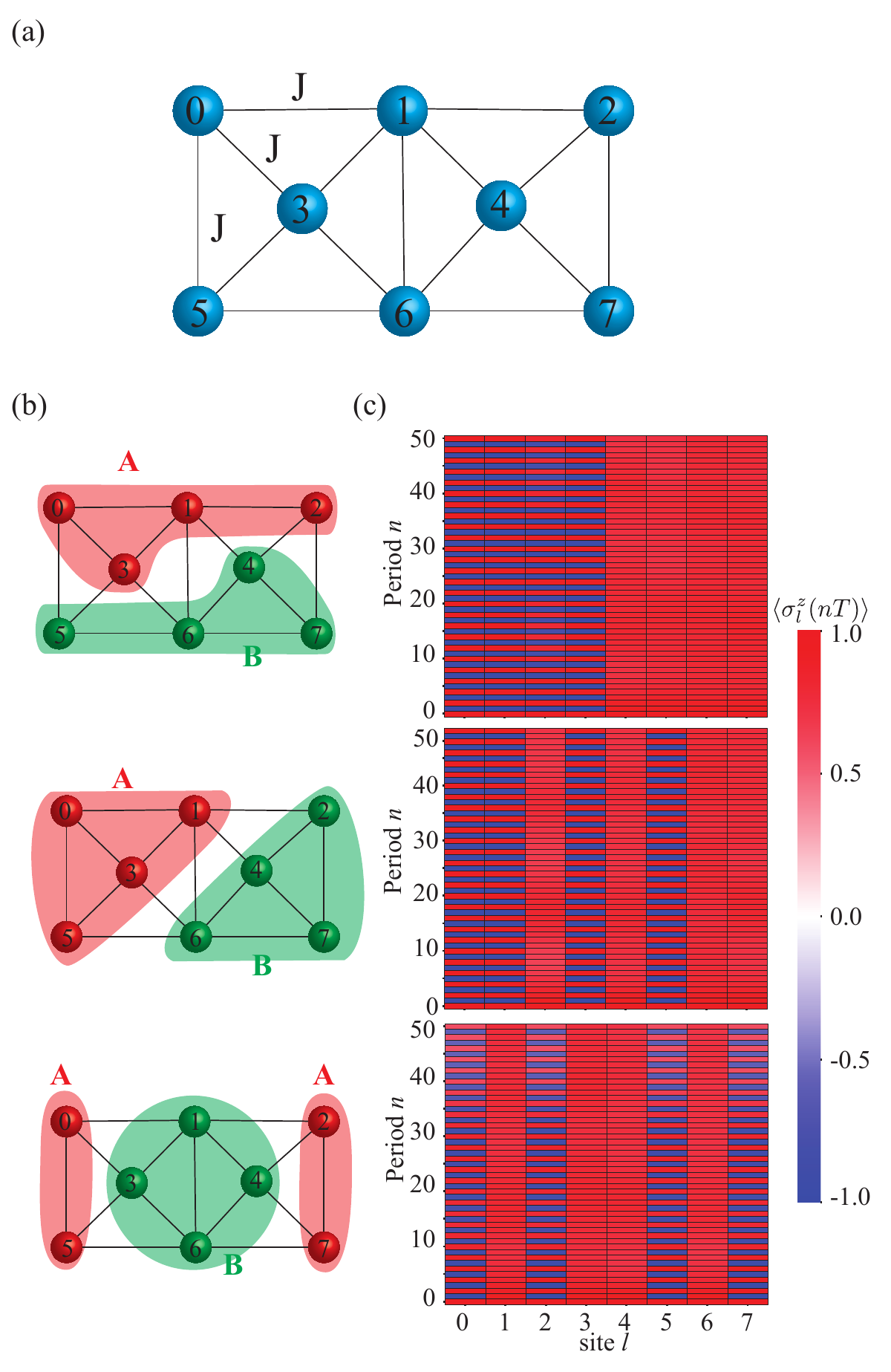}
\caption{Chimera DTC in a ladder-like quantum spin network. Panel (a) illustrates the connectivity of the spin network with $N=8$ nodes. The coupling between the spins is uniform with strength $JT = 0.2$. Correspondingly, (b) depicts different bipartitions of the network. Note that for certain partitions, the region A can have multiple connected components.  (c) shows the 
dynamics of ensemble averaged local magnetization $\langle\sigma_l^z(nT)\rangle$ for rotation errors $\epsilon_A =0.03$ and $\epsilon_B =0.9$. We calculate the ensemble average of the magnetization with 100 realizations of disorder. We set $gT= \pi$, $JT = 0.2$, and $WT=2\pi$. 
}
\label{Fig4S}
\end{figure*}

\section{
V. Chimera DTC in a ladder-like quantum spin network: The role of the bipartitions
\label{SecV}
} 
In our manuscript, we presented a general theory of Chimera DTCs in quantum spin networks. To substantiate our results, we considered a particular example with a one-dimensional arrangement of spins. In our example, the connectivity of the network was determined by a parameter that allows us to interpolate between different connectivities.

Here we provide additional numerical evidence for Chimera DTC in a network with a non-trivial connectivity, as depicted in Fig.~\ref{Fig4S}~a). We performed numerical calculations for $N=8$ spins on a network with a complex topology (a ladder-like structure). Fig.~\ref{Fig4S}~b) shows different kinds of bipartitions A and B of the network. For convenience, we consider the strong coupling regime $JT=0.2$ between the sites [see Fig.~\ref{Fig4S}~a)] and rotation errors $\epsilon_A=0.03$, $\epsilon_B= 0.9$ in regions A and B. We have averaged the local magnetization over $100$ realizations of disorder (with strength $WT=2\pi$). Further the initial state is fully polarized along the z axis. Remarkably, as one can see in the figure, the chimera nature of the dynamics appears for different partitions of the network, as one can see from the dynamics of the local magnetization $\langle\sigma_l^z(nT)\rangle$ in Fig.~\ref{Fig4S}~c).

\end{document}